# *Explainable Multimodal Machine Learning* for Revealing Structure-Property Relationships in Carbon Nanotube Fibers

D. Kimura, N. Tajima, T. Okazaki, S. Muroga*

Nano Carbon Device Research Center, National Institute of Advanced Industrial Science and Technology (AIST), Tsukuba Central 5, 1-1-1, Higashi, Tsukuba, Ibaraki, 305-8565 Japan

*Corresponding author: Shun Muroga, Nano Carbon Device Research Center, National Institute of Advanced Industrial Science and Technology (AIST), E-mail: muroga-sh@aist.go.jp

## Abstract

In this study, we propose *Explainable Multimodal Machine Learning* (EMML), which integrates the analysis of diverse data types (multimodal data) using factor analysis for feature extraction with *Explainable AI* (XAI), for carbon nanotube (CNT) fibers prepared from aqueous dispersions. This method is a powerful approach to elucidate the mechanisms governing material properties, where multi-stage fabrication conditions and multiscale structures have complex influences. Thus, in our case, this approach helps us understand how different processing steps and structures at various scales impact the final properties of CNT fibers. The analysis targeted structures ranging from the nanoscale to the macroscale, including aggregation size distributions of CNT dispersions and the effective length of CNTs. Furthermore, because some types of data were difficult to interpret using standard methods, challenging-to-interpret distribution data were analyzed using *Negative Matrix Factorization* (NMF) for extracting key features that determine the outcome. Contribution analysis with *SHapley Additive exPlanations* (SHAP) demonstrated that small, uniformly distributed aggregates are crucial for improving fracture strength, while CNTs with long effective lengths are significant factors for enhancing electrical conductivity. The analysis also identified thresholds and trends for these key factors to assist in defining the conditions needed to optimize CNT fiber properties. EMML is not limited to CNT fibers but can be applied to the design of other materials derived from nanomaterials, making it a useful tool for developing a wide range of advanced materials. This approach provides a foundation for advancing data-driven materials research.

## Highlight

- Proposed Explainable Multimodal Machine Learning for analyzing materials with multiscale structures.
- Demonstrated the utility of Negative Matrix Factorization for extracting interpretable features from complex distribution data.
- Identified key features influencing fracture strength, electrical conductivity, and Young's modulus, including thresholds and trends.

- Showed that small, uniformly distributed aggregates improve fracture strength, while long CNTs enhance electrical conductivity.
- Highlighted the applicability of EMML to other nanomaterial-based material designs, advancing data-driven materials research.

Graphical abstract

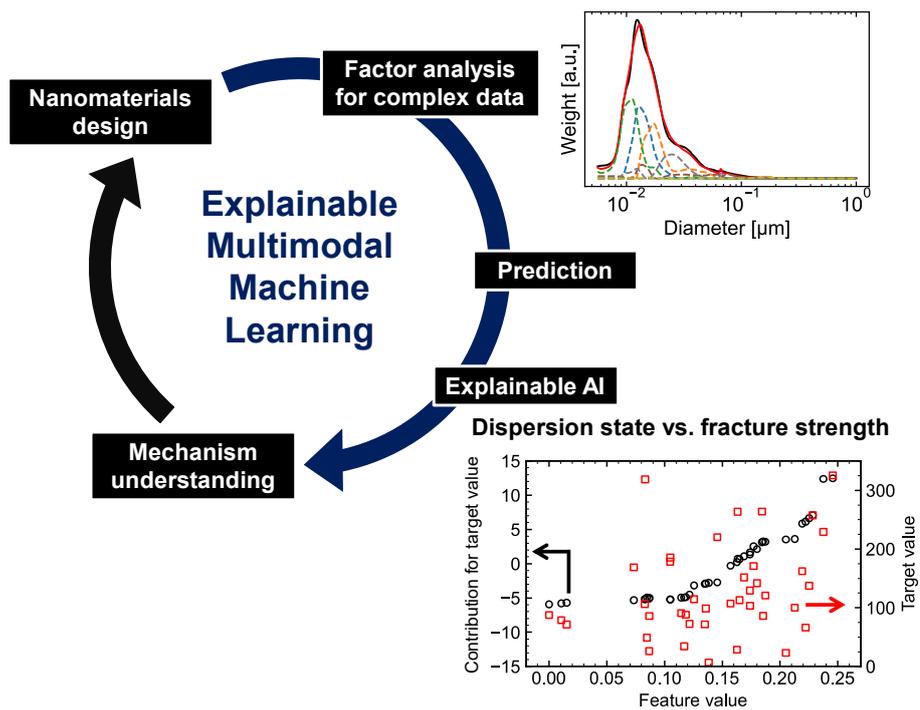

# 1. Introduction

Carbon nanotube (CNT) fibers [1–12] have garnered attention as next-generation materials due to their exceptional mechanical, electrical, and thermal properties. These fibers feature a structure in which CNTs, as nanoscale materials, assemble into bundles. There is great interest in transferring the remarkable properties of individual CNTs to CNT fibers [11–19]. Recent advancements in manufacturing processes have improved the properties of CNT fibers, with some reports indicating performance surpassing that of carbon fibers in terms of tensile strength and electrical conductivity [5]. Continued optimization of manufacturing conditions and elucidation of mechanisms governing material properties are expected to accelerate research and development toward innovative applications, such as space elevators and long-distance power lines [6, 7].

Conversely, bottom-up manufacturing processes for nanomaterials like CNT fibers require the optimization of multi-stage manufacturing conditions and multiscale structures ranging from the atomic to macroscopic scales [20–24]. These factors are intricately and mutually interconnected, and current technologies have yet to achieve sufficient optimization. Consequently, a gap remains between transfer of the intrinsic nanoscale properties of CNTs to the actual macroscopic performance of CNT fibers. This challenge is not unique to CNT fibers but is a common issue in materials research, representing one of the most critical barriers to next-generation material development. We recognize the rapid progress in the application of artificial intelligence (AI) in materials research in recent years and its capability to extract multidimensional and nonlinear patterns, making it a promising tool for addressing this complex challenges.

AI applications have also made significant progress in research related to CNTs. For instance, Y. Nonoguchi et al. introduced machine learning on molecular descriptors to determine optimal dispersion conditions for CNTs in organic solvents that are typically challenging for dispersion [25]. Similarly, H. Jintoku et al. successfully fabricated CNT transparent conducting films with unprecedented properties using a previously unknown dispersant predicted through experimental and molecular descriptors [26]. Furthermore, D. Lin et al. identified CNT synthesis conditions that try to optimize both crystallinity and length, two often mutually exclusive properties, through machine learning-based property prediction models and virtual experiments [27]. Recently, S. Muroga et al. pioneered a novel multimodal machine learning approach for predicting the properties of complex materials by integrating diverse data types [28]. The effectiveness of this method was demonstrated in CNT composites through the fusion of physical and chemical structural data, such as imaging and spectral information. This approach uncovered intricate correlations in material properties that were previously inaccessible through conventional single-modal analysis, highlighting the potential of artificial intelligence to advance materials research.

Previous studies on multimodal machine learning in materials research and materials informatics have predominantly focused on property prediction and exploration. However, studies focusing on model interpretability to explain the mechanisms behind machine learning predictions remain limited. In materials science, it is crucial to elucidate generalized mechanisms, rather than relying on models limited to specific experimental conditions, to provide innovative and accurate design guidelines. Moreover, few studies have attempted to elucidate mechanisms based on the interpretation of machine learning models for materials with complex relationships, such as those involving multi-stage manufacturing processes and multiscale structures, ranging from nanomaterials to functional materials. Some efforts have utilized *explainable AI* (XAI) to evaluate the importance of features in machine learning model predictions, demonstrating the potential of combining interpretable models with XAI as an effective tool for mechanism elucidation [29–31].

This study proposes *Explainable Multimodal Machine Learning* (EMML), which integrates multimodal machine learning and XAI for CNT fibers prepared from aqueous dispersions (Figure 1). The proposed method comprises multimodal machine learning, which combines multiscale analyses specialized for CNT fiber characterization (e.g., effective length and aggregate size distribution in dispersions), factor analysis, and regression analysis, along with XAI, using *SHapley Additive*

*exPlanations* (SHAP) [32] to provide global and local interpretations of prediction results. EMML offers a comprehensive approach to address the complexity of multi-stage manufacturing conditions and multiscale structures, aiming to optimize material properties and elucidate mechanisms governing material properties. This methodology holds significance as a foundational technology to facilitate the practical application of bottom-up-designed functional materials composed of nanomaterials

## 2. Material and methods

The dataset in this study is derived from CNT fibers prepared from aqueous dispersions reported by Tajima et al. [10]. The preparation of CNT fibers using aqueous dispersions has garnered attention as a highly safe and sustainable method, with several studies having been reported. The dispersions used in this study were prepared using single-walled CNTs synthesized via the eDIPS method, water, and either sodium cholate or sodium taurodeoxycholate as dispersants. The dispersion process employed a homogenizer, a milling mixer, and a Nanovater.

To evaluate the multiscale structures, the following methods were utilized. Additionally, the properties of CNT fibers, including fracture strength, electrical conductivity, and Young's modulus, were measured and used as target variables in the regression analysis.

- Disc Centrifuge (DCS): CNT aggregation size distribution [33]
- Far-infrared Spectroscopy (FIR): Effective CNT length [34]
- Raman Spectroscopy: Extent of hexagonal carbon networks and defect structures [35, 36]
- Morphological Measurements: Density, linear density, cross-sectional area, and void fraction

Notably, DCS and FIR are specialized analytical techniques for CNTs and are highly effective in understanding the multiscale structures of dispersions and fibers.

Distribution data obtained from DCS often makes it challenging to identify which aspects of the data characterize the material properties. Furthermore, using such data directly in regression analysis can lead to models with limited interpretability. To address this issue, this study compared *Principal Component Analysis* (PCA) and *Negative Matrix Factorization* (NMF) [37] for feature extraction from distribution data. PCA is one of the most fundamental and widely used dimensionality reduction techniques, which efficiently projects high-dimensional data onto a lower-dimensional space based on orthogonal bases that maximizes data variance. In contrast, NMF is a method that decomposes data into non-negative components. For non-negative data, such as distribution data, it generates highly sparse bases, making it a promising approach for extracting interpretable features (Figure S1).

NMF was used to extract features from DCS data after comparing its performance with PCA. These features were integrated with other analytical results and fabrication conditions to perform regression analysis on the properties of CNT fibers. A Random Forest model was adopted for regression, and the coefficient of determination was calculated using *Leave-One-Out Cross-Validation* (LOOCV) to evaluate predictive performance. Furthermore, the number of factors in NMF was optimized based on predictive performance (Figure S2).

Finally, SHAP values were calculated for the model trained under optimized conditions. SHAP is one of the representative methods of XAI and is used to visualize the contributions of explanatory variables to predictions of the target variable. A global evaluation was conducted based on the magnitude of SHAP values for each explanatory variable to assess their importance. Additionally, a local evaluation was performed based on the changes in SHAP interaction values for individual samples, thereby elucidating the relationships between explanatory variables and the target variable (Figure S3).

XAI does not directly clarify mechanisms, but it is expected to contribute to elucidating material mechanisms through interpretable learning models combined with researchers' domain knowledge.

# 3. Results and discussion

3.1 Multiscale Structural Analysis Techniques and Factor Analysis of Distribution Data

CNT fibers exhibit multiscale structures, with features ranging from nanoscale defects and aggregation structures to macroscale formations. To accurately investigate these hierarchical structures, advanced analytical techniques specialized for CNTs have been developed. These techniques are crucial for maximizing the exceptional properties of CNT fibers in functional material development. While they provide detailed insights into nanomaterial states, a major challenge lies in the interpretation of complex data. In particular, for distribution and spectral data with ambiguous peak structures, it is critical to extract features that strongly influence material properties.

Figure 2 presents the results of applying PCA and NMF to CNT aggregation size distributions of 40 types of dispersions. For distributions with a single dominant peak, shown in Figure 2(a, c), both PCA and NMF successfully reconstructed the original data. However, for multi-peaked distributions, shown in Figure 2(b, d), PCA failed to adequately reproduce the distribution shape. Furthermore, for the single-peaked distributions in Figure 2(a, c), the second and third principal components of PCA exhibited wave-like distributions, which lack clear physical interpretation. In contrast, NMF decomposed the data into three distributions with distinct peak positions, providing highly interpretable results. These findings indicate that NMF is more suitable for factor analysis of distribution data when interpretability is a priority.

3.2 Factor Analysis and Feature Extraction Using NMF

Figure 3(a) shows the results of regression analysis performed using features extracted by NMF with the number of components varying from 1 to 40. For fracture strength, electrical conductivity, and Young's modulus, the coefficient of determination reached its maximum when the number of components was set to 7. This indicates that using 7 key factors provided the best explanation of how these properties are influenced by the underlying structure and processing conditions. As shown in Figures 3(b-1), (b-2), and (b-3), the predicted and experimental values for the test data closely matched the experimental values, demonstrating the reliability of the regression model. Furthermore, beyond 7 components, predictive accuracy declined, as shown in Figure 3(a). This decline is likely due to the inclusion of components unrelated to material properties or noise, which increased data sparsity and hindered pattern recognition.

Figure 4(a) illustrates the classification results of the basis distributions obtained through NMF. The basis distributions are arranged vertically according to the primary aggregation size, determined by the main peak position, and are further grouped horizontally based on peak characteristics. The seven classified basis distributions exhibited high interpretability and could be broadly categorized into three groups: small, medium, and large sizes. The small-size group comprised of four components, the medium-size group two components, and the large-size group one component. These results suggest that the properties of CNT fibers are influenced by the aggregation size and distribution width of the dispersions. Figure 4(b) further illustrates how the aggregation size distributions obtained from DCS were represented using the basis distributions. Unlike X-ray diffraction data, which have theoretical bases for peak positions and peak functions, distribution data lack such theoretical foundations, making them difficult to analyze using conventional fitting methods. However, it was confirmed that interpretable basis distributions could be reconstructed by employing NMF. Thus, NMF has proven useful for analyzing multi-peaked and nonlinear data.

## 3.3 Global and Local Analyses Using XAI

Figure 5 provides a global evaluation of how explanatory variables to regression predictions. The vertical axis lists the explanatory variables, ranked by their variance, with higher variance variables ranked at the top. Among these, density and linear density exhibited the strongest contributions to CNT fiber properties. However, in importance analysis of explanatory variables, scaling factors among the top-ranked variables tended to be overestimated, suggesting that variables ranked third and beyond may also have significant impacts. For example, fracture strength and electrical conductivity were notably influenced by aggregation size features extracted from DCS and effective CNT length from FIR. These findings indicate that the dispersion state and nanoscale structure of CNTs influence both fracture strength and electrical conductivity. In contrast, Young's modulus was more strongly influenced by density than other properties, while the effects of nanoscale structures from DCS and FIR were relatively minor. This suggests that Young's modulus is strongly dependent on macroscale morphology, particularly on density, void structures, and fiber diameter.

Figures 9–11 illustrate the local relationships between explanatory and target variables. The horizontal axis represents explanatory variables, while the left vertical axis (●) shows their contributions, and the right vertical axis (▲) shows the target variable values. Figure 6 confirms that macroscale structures, such as fiber density and linear density, influence fracture strength, electrical conductivity, and Young's modulus in similar ways. Higher density and smaller fiber diameter were found to improve material properties.

In Figures 10 and 11, SHAP interaction values (left vertical axis) are scaled to approximately ±5% of the maximum target variable value (right vertical axis), including fracture strength, electrical conductivity, and Young's modulus. As previously noted in Figure 5, top-ranked explanatory variables tend to be overestimated as scaling factors, potentially distorting their actual contributions. Consequently, a threshold of 10% is considered meaningful contribution for interpretation. In Figure 7, a threshold contribution of $I_G/I_D$ was observed ~30 for fracture strength and electrical conductivity, suggesting that the $I_G/I_D$ ratio plays a critical role in mechanical interactions and the formation of conduction paths. Additionally, when the effective length of CNTs exceeded 2000 nm, electrical conductivity improved significantly, while its effect on fracture strength remained minimal. This highlights the importance of CNT continuity and the absence of kinks for electrical conductivity.

Figure 8 explores the relationship between DCS-derived aggregation size features and target variables. The aggregation size features (DCS_Basis000, DCS_Basis001, …) indicate that larger values correspond to higher weights of the respective basis distributions in each sample's aggregation size distribution. For fracture strength, small and uniformly distributed aggregates (Basis000, Basis001) and a small amount of medium-sized aggregates with wide distribution widths (Basis004) were found to be beneficial. Conversely, even a slight presence of large aggregates (Basis006) resulted in a significant decrease in fracture strength. For electrical conductivity, large aggregates (Basis006) showed a less pronounced effect compared to fracture strength. However, high conductivity required a narrow distribution of small aggregates, meaning that the broadening component (Basis001) should be minimized. Thus, the dispersion method of CNTs and the state of the dispersion are central in creating this narrow distribution to realize the enhancement in electrical performance of CNT fibers. Previous studies have also reported importance of the dispersion state on the spinning processes [38], which highlights the necessity of accurately evaluating CNT dispersions. Overall, this study highlights that both macroscale and nanoscale structures significantly impact CNT fiber properties. Density and linear density are primary determinants of material performance, with dispersion state and CNT length playing crucial roles in electrical conductivity and fracture strength. Young's modulus is predominantly influenced by macroscale factors such as density and fiber morphology. Taken together, these findings emphasize the importance of accurately characterizing CNT dispersion and structure to optimize fiber properties, reinforcing previous research conclusions on the necessity of proper CNT processing techniques.

## 3.4 Mechanism Elucidation of CNT Fibers Using EMML

Figure 9 summarizes the findings and highlights the effectiveness of EMML in clarifying the relationships between the structures with the material properties in CNT fibers. In short, high Young's modulus requires high packing density resulting in high density and low void content. Practically, this stems from the spinning conditions. In contrast, high fracture strength requires dispersions with small, uniformly distributed aggregates and a small amount of medium-sized aggregates with wide distribution widths. For electrical conductivity, maintaining CNTs with fewer defects is found to be important as evidenced by the need for longer effective lengths as well as a narrow distribution of small aggregates. Furthermore, for Young's modulus, the results suggested that macroscale structures, shaped by spinning conditions, exert a stronger influence than the aggregation characteristics of the dispersions.

The structure of CNT fibers follows a hierarchical arrangement in which individual CNTs form bundles, and these bundles collectively assemble into fibers. We interpret that within this framework, the structural organization of CNT aggregates, rather than individual CNTs, serves as the fundamental building block in determining material properties. Given that primary processing objectives are either load transfer or electron transport, the number and intimacy of contacts between aggregates are key, which depend on their size and distribution. Since Young's modulus characterizes the material response in the small strain region during tensile deformation, it is predominantly influenced by the higher-order structure formed by the CNT bundles. As a result, density and fiber morphology are the dominant factors. In contrast, fracture strength, which is the stress level at which a material fails, is governed by the local structure surrounding defects, as described by the weakest-link model [39]. It is highly dependent on the internal structural evolution and crack propagation during deformation. Thus, the interactions between CNTs, including entanglements both within and between bundles, play a crucial role in determining fracture strength. This explains the importance of a narrow distribution of small aggregates as well as the $I_G/I_D$-ratio in enhancing fracture strength. A narrow distribution of small particle maximizes the number of aggregate-aggregate junctions to provide load transfer and avoid stress concentration. The significance of the $I_G/I_D$-ratio lies in its role as a measure of graphitic crystallinity and carbonaceous impurities; the latter can disrupt intimate contacts and weaken fiber integrity. In addition, our results show that a small fraction of medium-sized aggregates also are a key determinant in fracture strength. We interpret that the presence of these aggregates introduces subtle structural heterogeneity, which helps to reduce crack and promote the redistribution of stress. We note that the effective length, a measure of CNT crystallinity alone, would be expected to play a critical role in fracture strength, but our analysis suggests that other structural factors currently limit fiber performance within our dataset. Finally, electrical conductivity follows expected trends and aligns with previous findings, reinforcing the idea that aggregate packing and CNT crystallinity remain primary determinants of this property. This is consistent with our results which highlight the high feature importance of a narrow distribution of small aggregates and long effective lengths in reducing internal resistance by maximizing the number and quality of electron pathways. Overall, these observations emphasize the need to consider aggregate assembly as the fundamental unit in CNT fiber processing. Future studies should further explore the specific roles of CNT crystallinity and length in mechanical failure to refine our understanding of these relationships.

Specifically, the bulk structure of CNT fibers fabricated from aqueous dispersions is primarily determined by macroscale characteristics, such as assembly (density). In contrast, the local structures that influence fracture behavior are shaped by sub-microscale features derived from aggregation structures in the dispersions, as well as nanoscale characteristics such as the $I_G/I_D$-ratio. These findings suggest that in the current CNT fibers fabricated from aqueous dispersions, the nanoscale structures and properties of CNTs are not fully translated in the bulk fiber structure. Consequently, this mismatch leads to inhomogeneous stress distribution and localized stress concentration within the fiber cross-section. Such structural inconsistencies prevent the fiber from fully utilizing the superior mechanical properties of CNTs, ultimately limiting its performance.

This study demonstrated the effectiveness of EMML, which integrates multiscale structural analysis, factor analysis, and XAI, in evaluating CNT fibers prepared from aqueous dispersions. The approach not only clarifies the optimization processes and structures for each property but also reveals critical thresholds and trends. By integrating physical models and simulations, the findings of this study are expected to validate the proposed mechanisms further and contribute to a deeper understanding of CNT fiber behavior.

## 4. Conclusion

In conclusion, we applied EMML (Explainable Machine Learning) to analyze CNT fibers fabricated from aqueous dispersions, aiming to elucidate the complex mechanisms governing their material properties under multi-stage fabrication conditions and multiscale structural influences. EMML integrated multimodal data analysis through factor analysis and enhances interpretability using Explainable AI (XAI). Through this approach, we identified the key factors influencing fracture strength, electrical conductivity, and Young's modulus. Fracture strength was found to be dependent on small, uniformly distributed aggregates and a small amount of medium-sized aggregates with wide distribution widths, which promote efficient load transfer while minimizing stress concentration. Electrical conductivity was primarily influenced by CNTs with fewer defects and longer effective lengths, ensuring uninterrupted electron transport pathways. In contrast, Young's modulus, a measure of a material's stiffness, was more significantly affected by the final macroscale fiber structure than by dispersion characteristics, highlighting the dominant role of fiber packing density and morphology in determining elastic response. The identification and understanding of the key factors governing these properties highlight the competing requirements of each property and provide insight into the necessary processing steps for optimizing fiber performance in a given application within this multi-stage fabrication process. Future integration with physical models and simulations is expected to further refine these findings and deepen the understanding of the underlying mechanisms. While conventional machine learning has successfully identified the structural importance of simple materials in a qualitative manner, it has often fallen short in providing sufficient insight into the mechanisms governing more complex materials. Our study using EMML demonstrated its effectiveness in identifying key factors, thresholds, and trends in the multi-stage fabrication process of CNT fibers, providing a robust framework for optimizing complex material design and manufacturing.

Acknowledgments

We appreciate Dr. Don N. Futaba for his advice about this paper.

CRediT authorship contribution statement

Conceptualization: S.M., T.O., Methodology: D. K. S.M., Software: D. K., S.M., Formal Analysis: D. K., S.M., Investigation: D. K., N. T., T. O., S.M., Visualization: D. K., S.M., Project Administration: S. M., Supervision: S.M., Writing – Original Draft: D.K., Writing – Review&Editing: S. M.

ORCID

Daisuke Kimura https://orcid.org/0009-0004-1182-7051
Toshiya Okazaki https://orcid.org/0000-0002-5958-0148
Shun Muroga https://orcid.org/0000-0002-6330-0436

# Figures

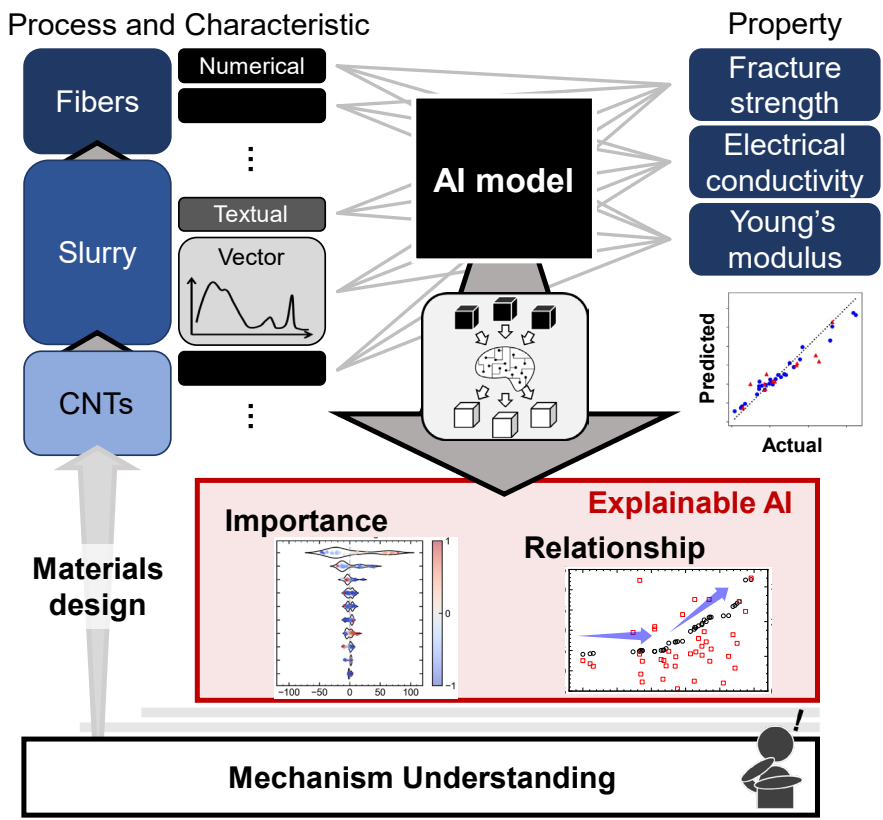

Figure 1. Overview of property predictions utilizing machine learning and explainable AI to support mechanism understanding.

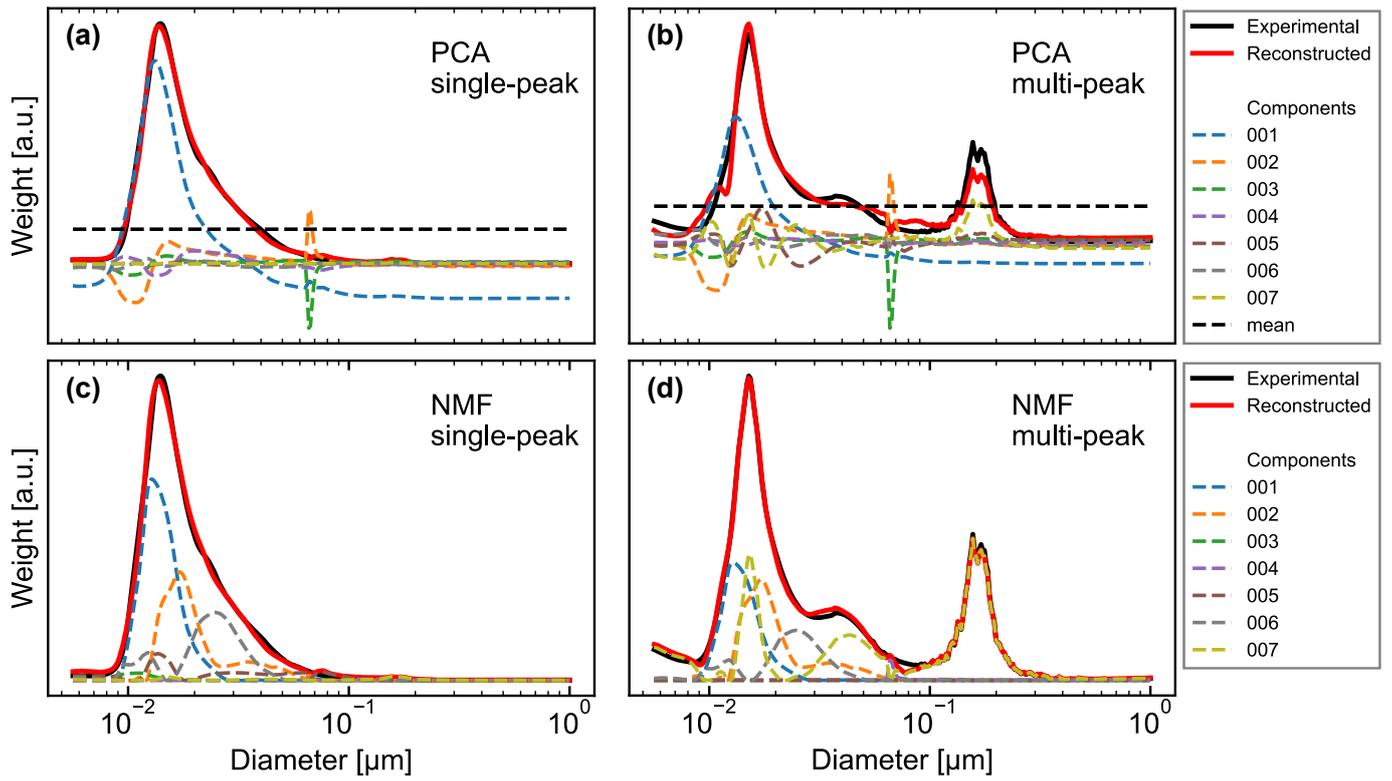

Figure 2. Results of decomposing (a, c) single-peaked and (b, d) multi-peaked distributions into 7 components using PCA and NMF.

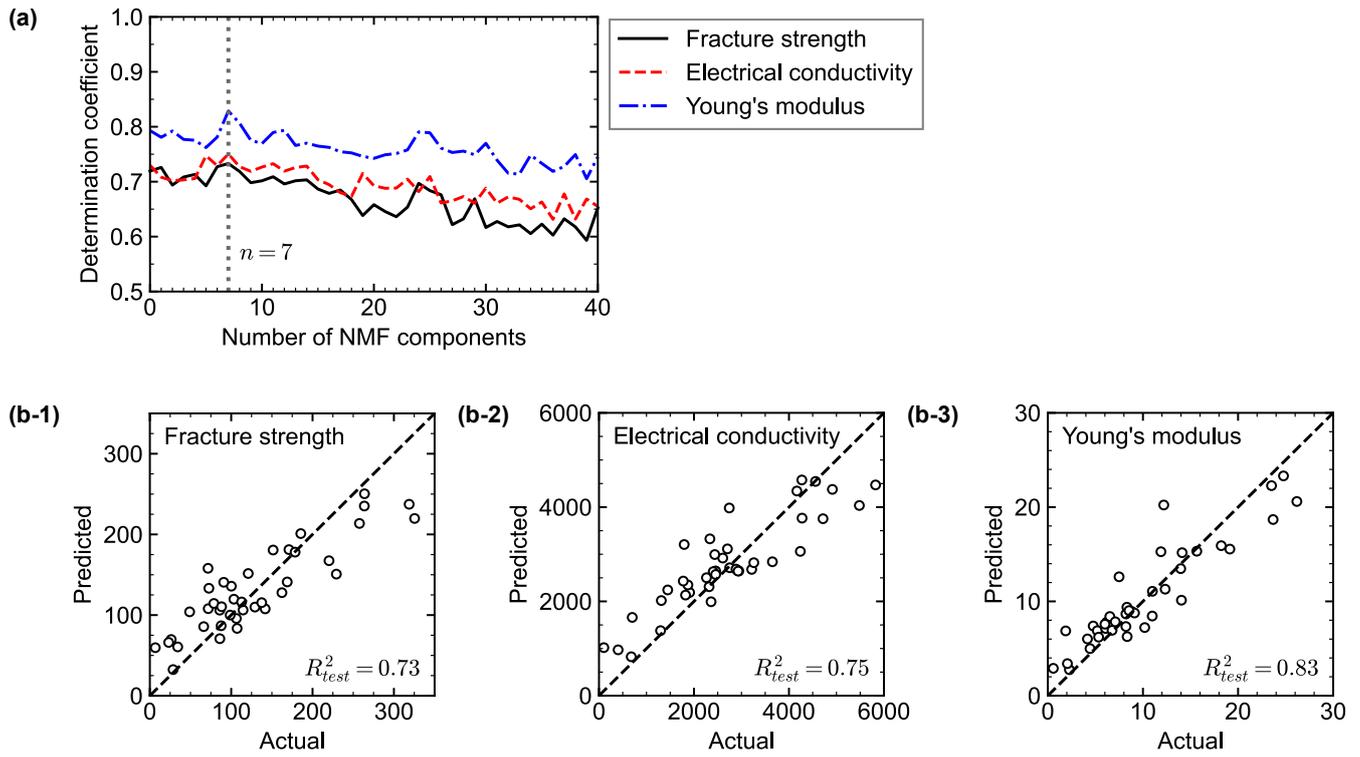

Figure 3. (a) Relationship between the number of NMF components and the coefficient of determination ($R^2$). At the optimal component number of 7, the relationships between actual and predicted values for (b-1) fracture strength, (b-2) electrical conductivity, and (b-3) Young's modulus are shown.

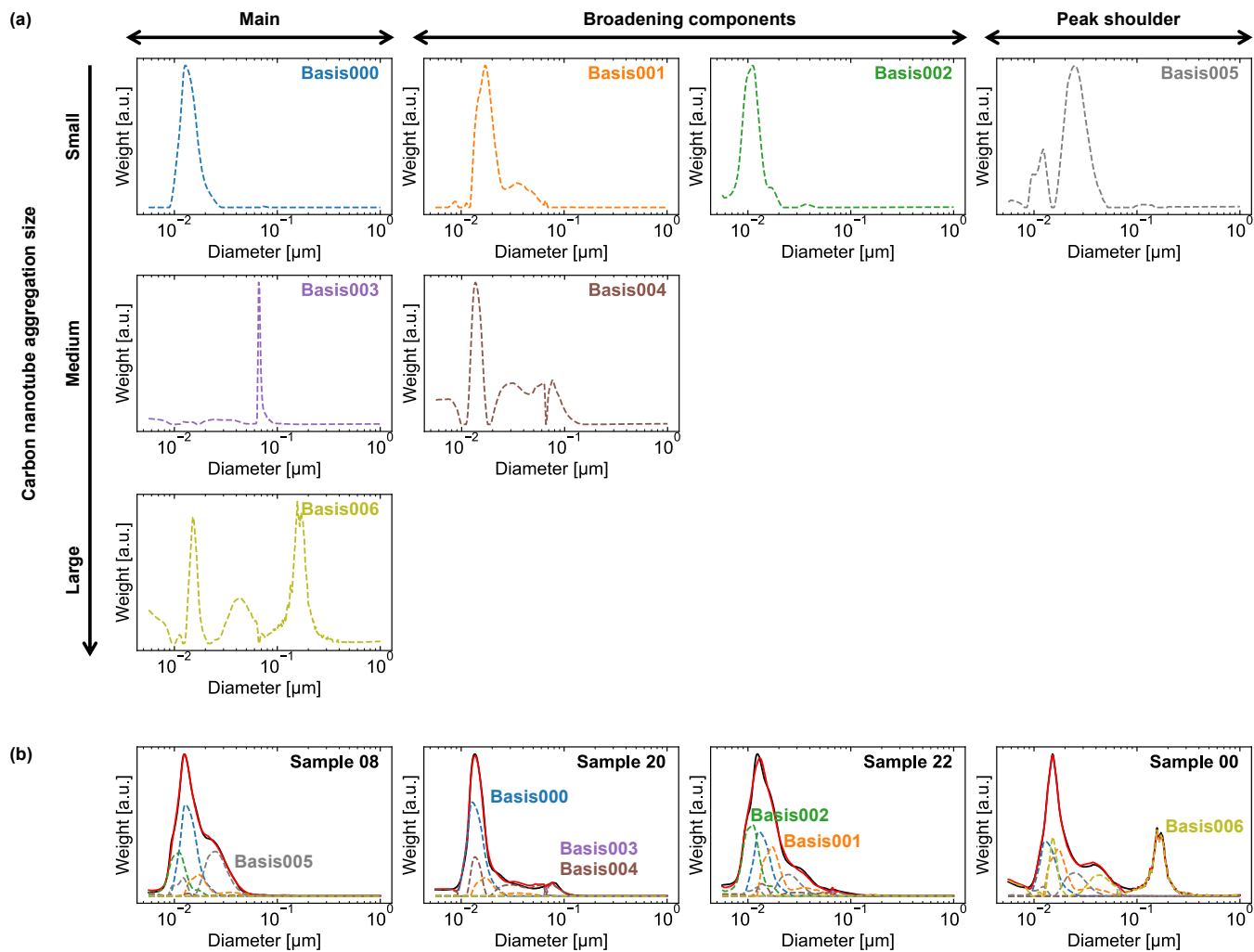

Figure 4. (a) Basis distributions and their classification in the aggregation size distribution of CNT dispersions. Small sizes are explained by four factors (Basis000, Basis001, Basis002, Basis005), medium sizes by two factors (Basis003, Basis004), and large sizes by one factor (Basis006). (b) Relationships between each basis distribution and the aggregation size distributions of representative samples.

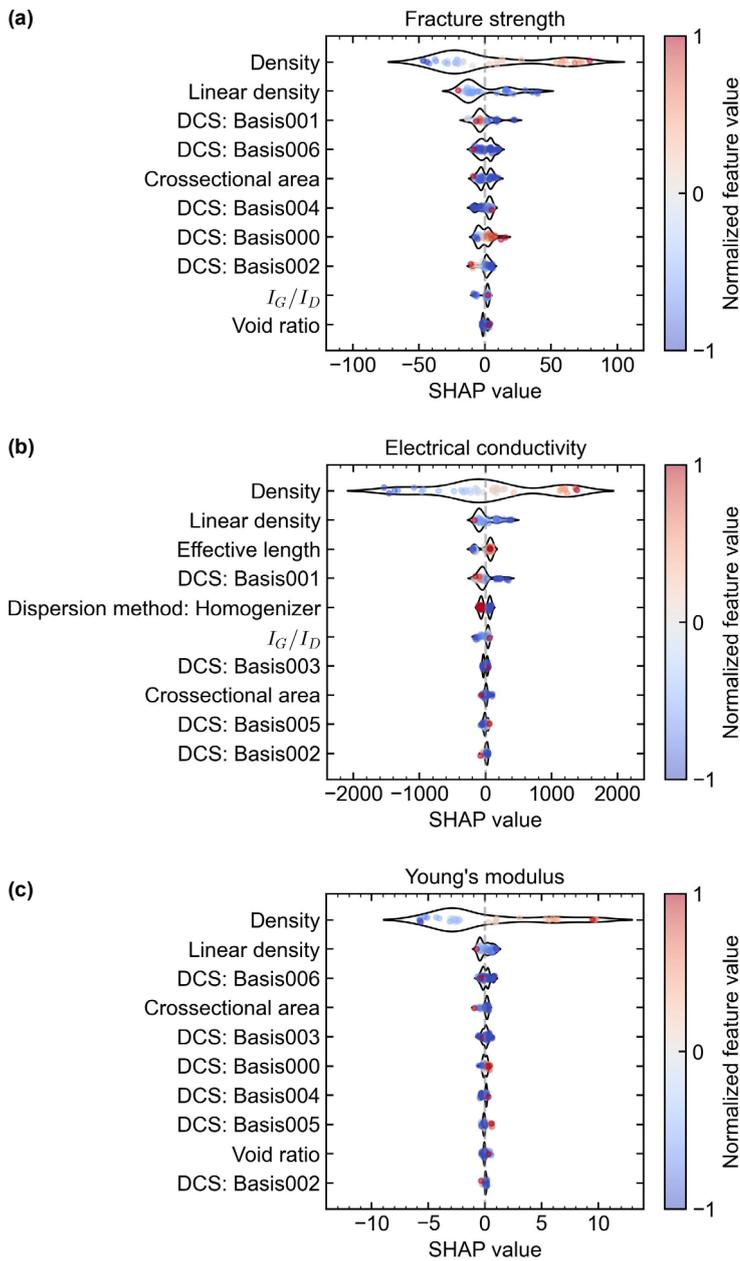

Figure 5. SHAP values for (a) fracture strength, (b) electrical conductivity, and (c) Young's modulus. The left axis indicates the top 10 explanatory variables, sorted in descending order based on feature importance. The plot's color and the color bar on the right represent the normalized magnitude of each explanatory variable, while the horizontal axis represents the SHAP values.

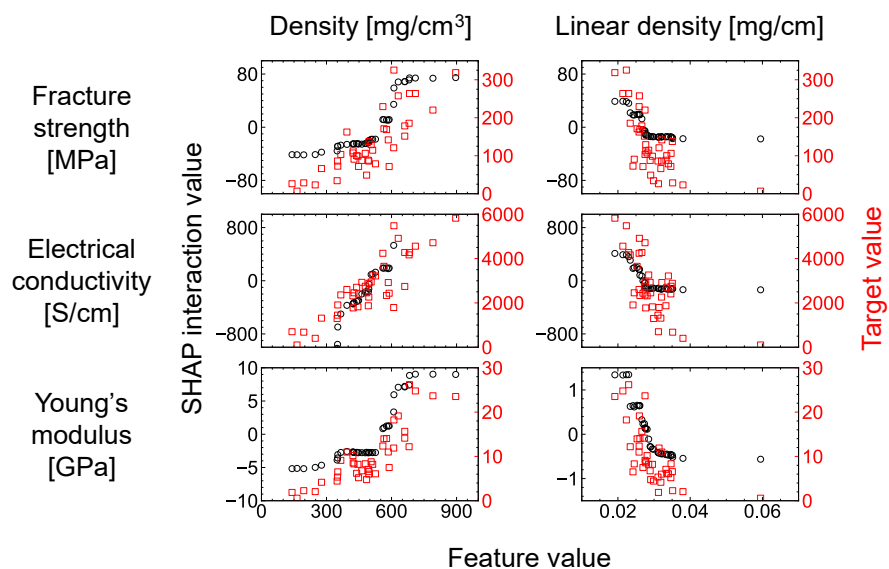

Figure 6. Relationship between explanatory variables related to the microscale morphology of fibers, their contributions, and the target variable. The horizontal axis represents the explanatory variables, the left vertical axis represents the SHAP values of the explanatory variables with respect to the target variable, and the right vertical axis indicates the target variable.

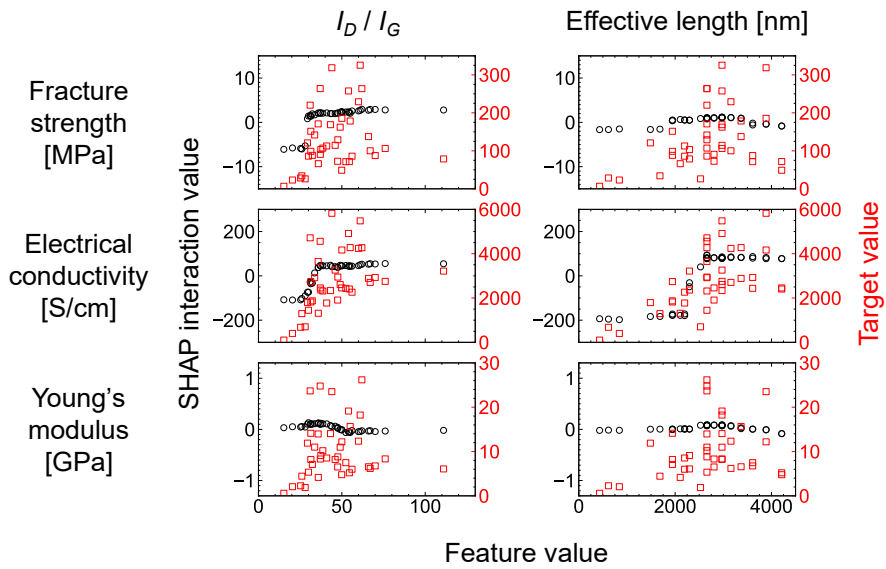

Figure 7. Relationship between explanatory variables related to the nanoscale structure of fibers, their contributions, and the target variable. The horizontal axis represents the explanatory variables, the left vertical axis represents the SHAP values of the explanatory variables with respect to the target variable, and the right vertical axis represents the target variable. The range of the left vertical axis is fixed to approximately 10% of the maximum value of the target variable.

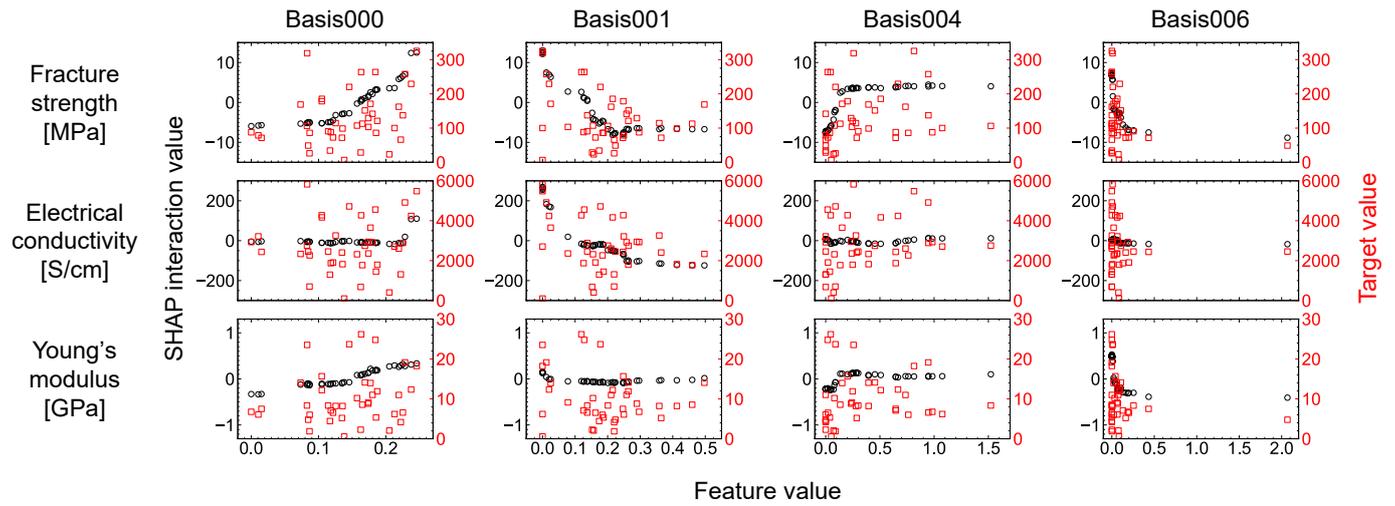

Figure 8. Relationship between the factor weights of the aggregation size distribution of CNT dispersions, their contributions, and the target variable. The horizontal axis represents the explanatory variables, the left vertical axis represents the SHAP values of the explanatory variables with respect to the target variable, and the right vertical axis represents the target variable. The range of the left vertical axis is fixed to approximately 10% of the maximum value of the target variable.

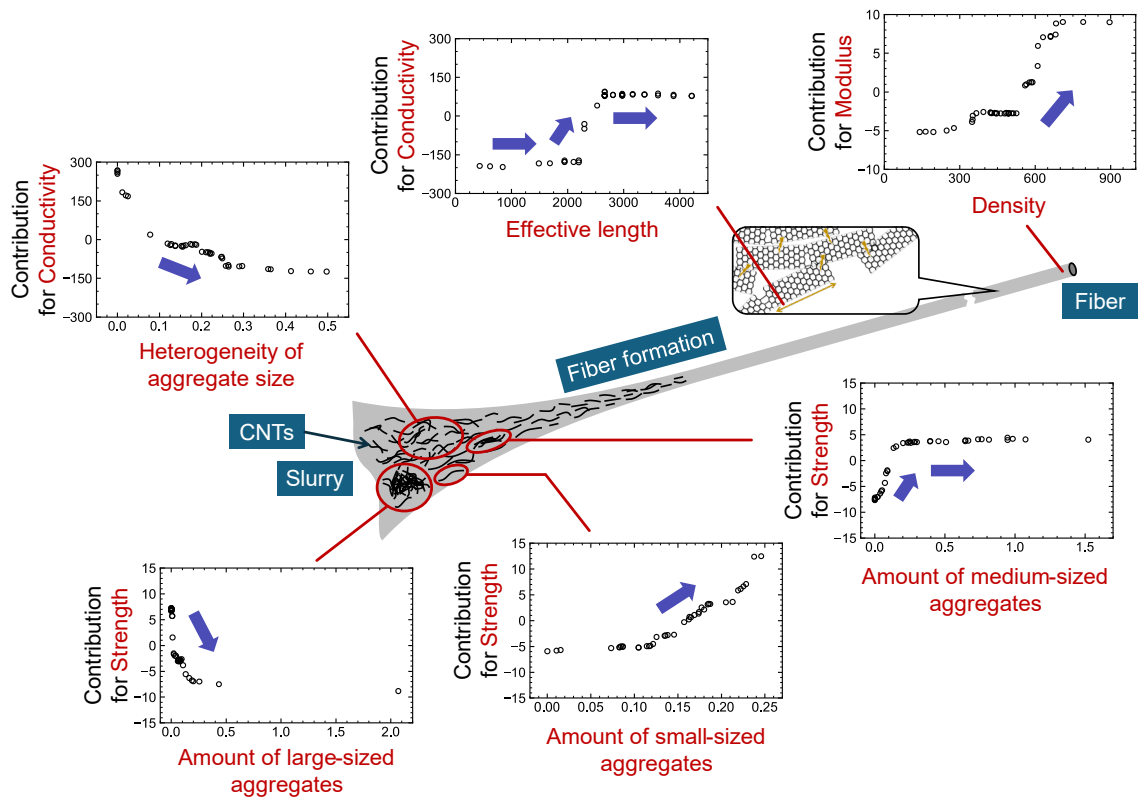

Figure 9. Relationship between the structure and properties of CNT fibers from aqueous dispersion. The multiscale structures of the fibers and the spinning dopes influence fracture strength, electrical conductivity, and Young's modulus.

Supplementary information

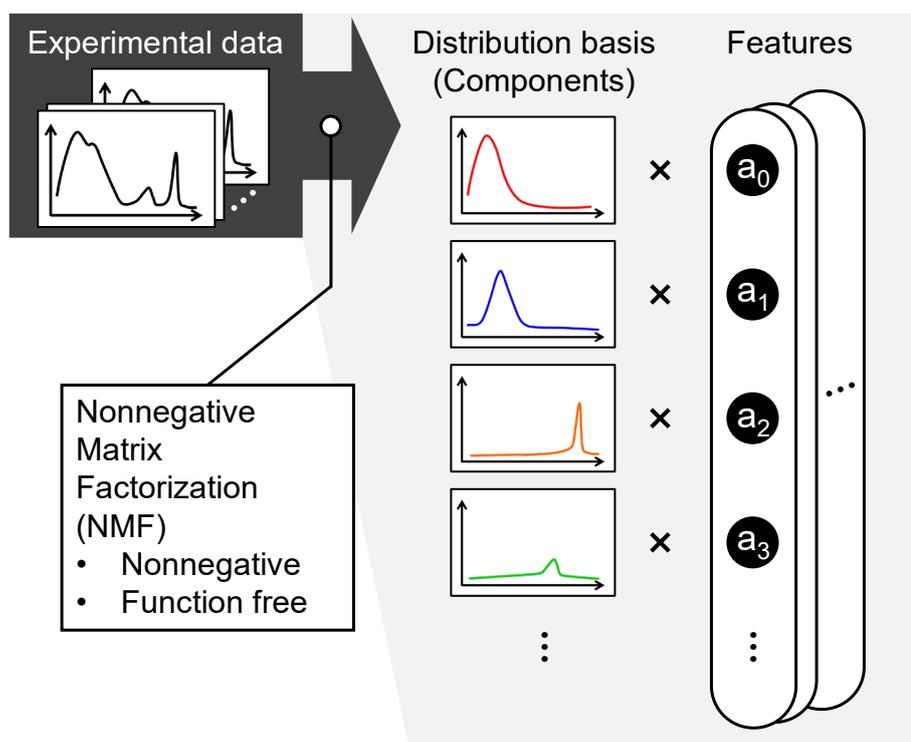

Figure S1. Schematic diagram of the decomposition of CNT dispersion aggregation size distribution data obtained from disc centrifugation using Nonnegative Matrix Factorization (NMF). Experimental data were decomposed into components and feature values, which were then utilized for predicting target variables.

**(a) Machine learning prediction**

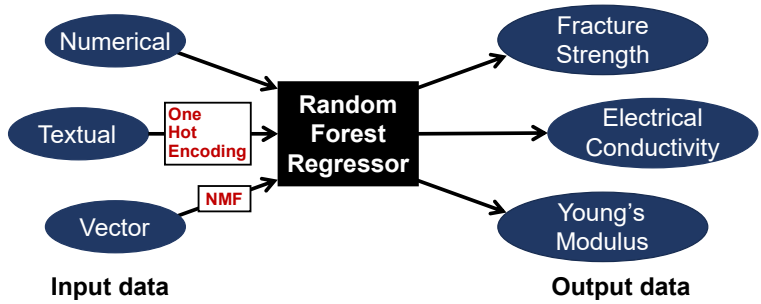

**(b) Leave-One-Out Cross-Validation (LOOCV)**

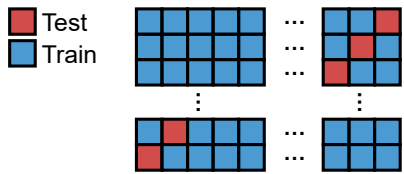

**(c) Hyperparameter tuning**

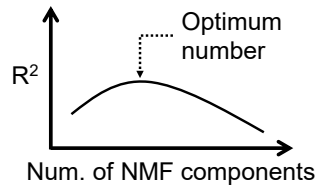

Figure S2. Procedure of the calculation process conducted in this study. (a) Text data were encoded using One-Hot Encoding, and vector data representing aggregation size distributions were transformed into feature values via NMF, followed by Random Forest regression. (b) The trained models were evaluated using Leave-One-Out cross-validation. (c) The optimal hyperparameters were determined based on the highest coefficient of determination ($R^2$).

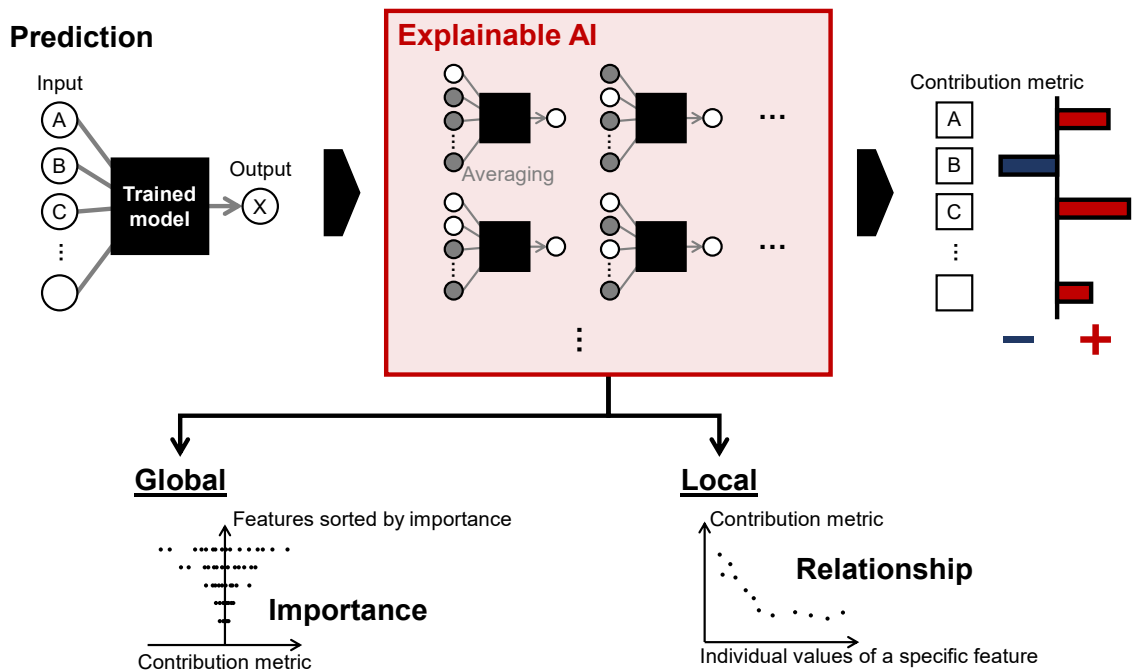

Figure S3. Illustration of the SHAP value calculation method and the graphical overview derived from the results. SHAP values provide a global evaluation of feature importance while also uncovering local relationships by focusing on changes in SHAP values for individual samples, thereby revealing the contributions of explanatory variables to the target variable.

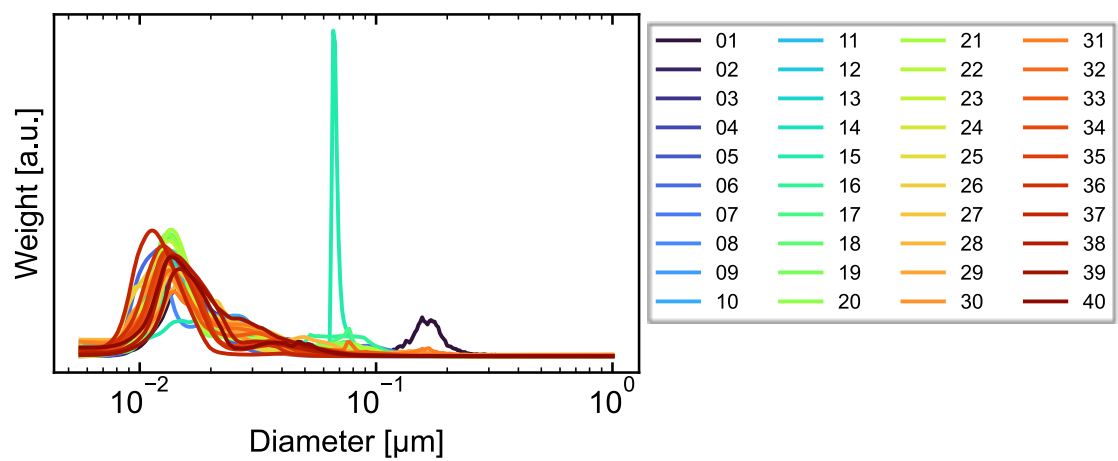

Figure S4. Aggregation size distribution measured by DCS for each sample.

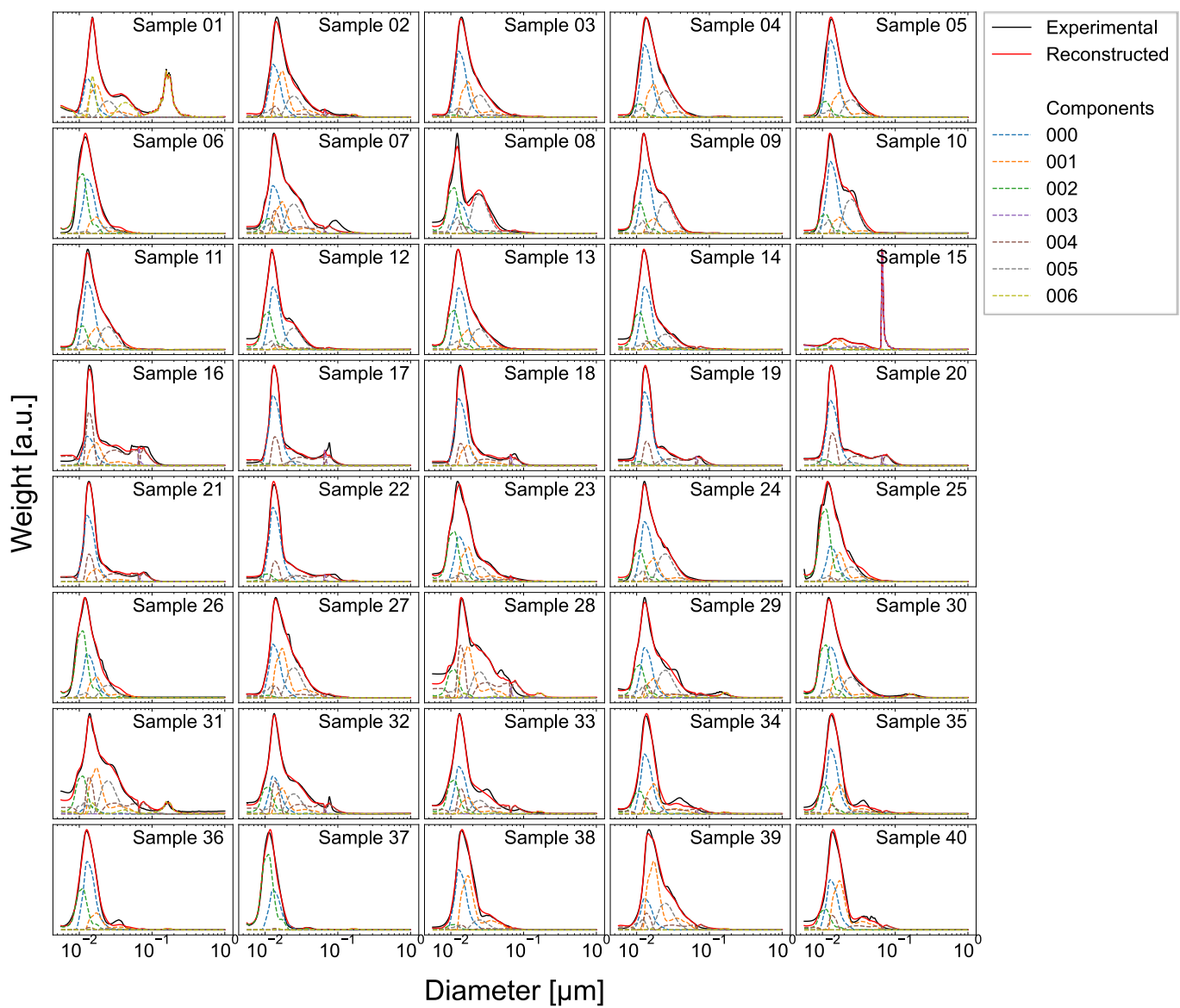

Figure S5. Relationship between the basis distributions obtained by NMF and the aggregation size distributions for all samples.

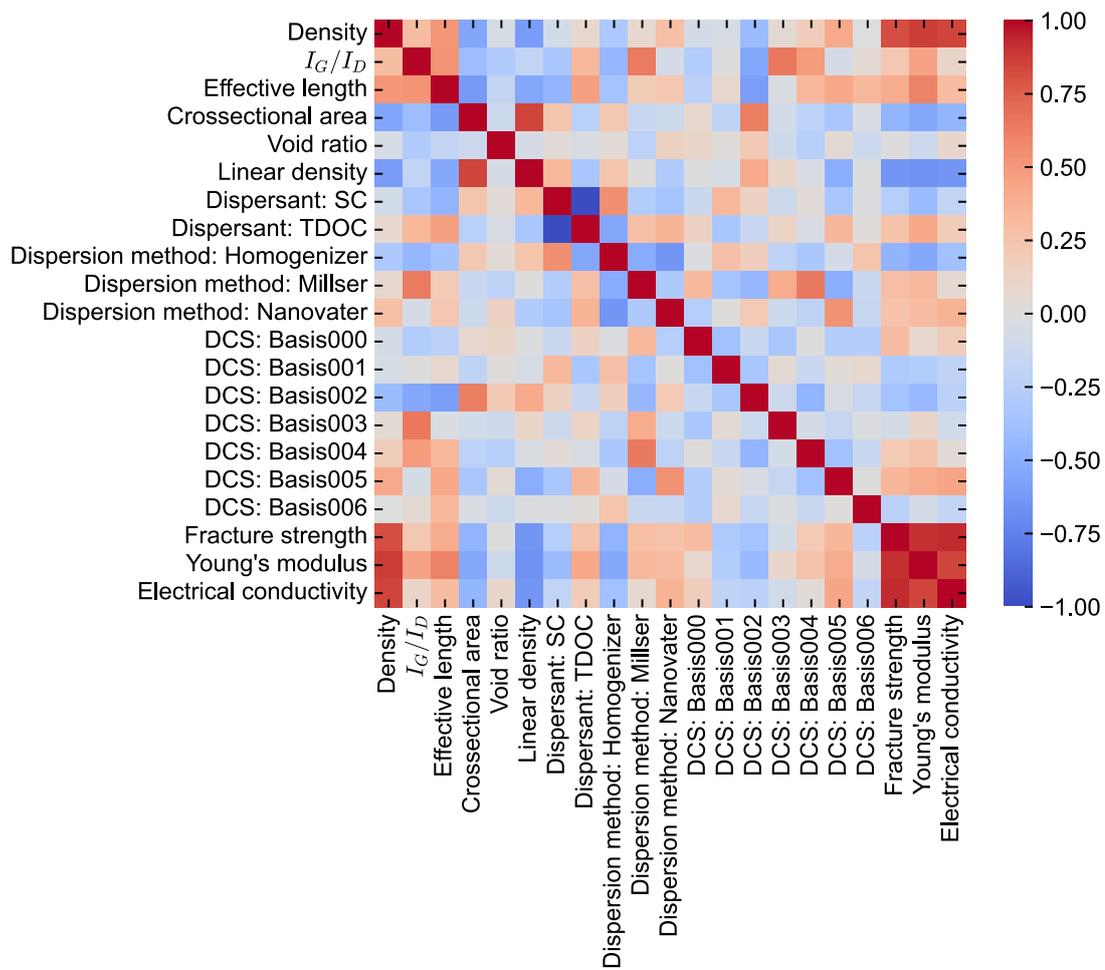

Figure S6. Heatmap of the correlation matrix between explanatory variables and the target variables.

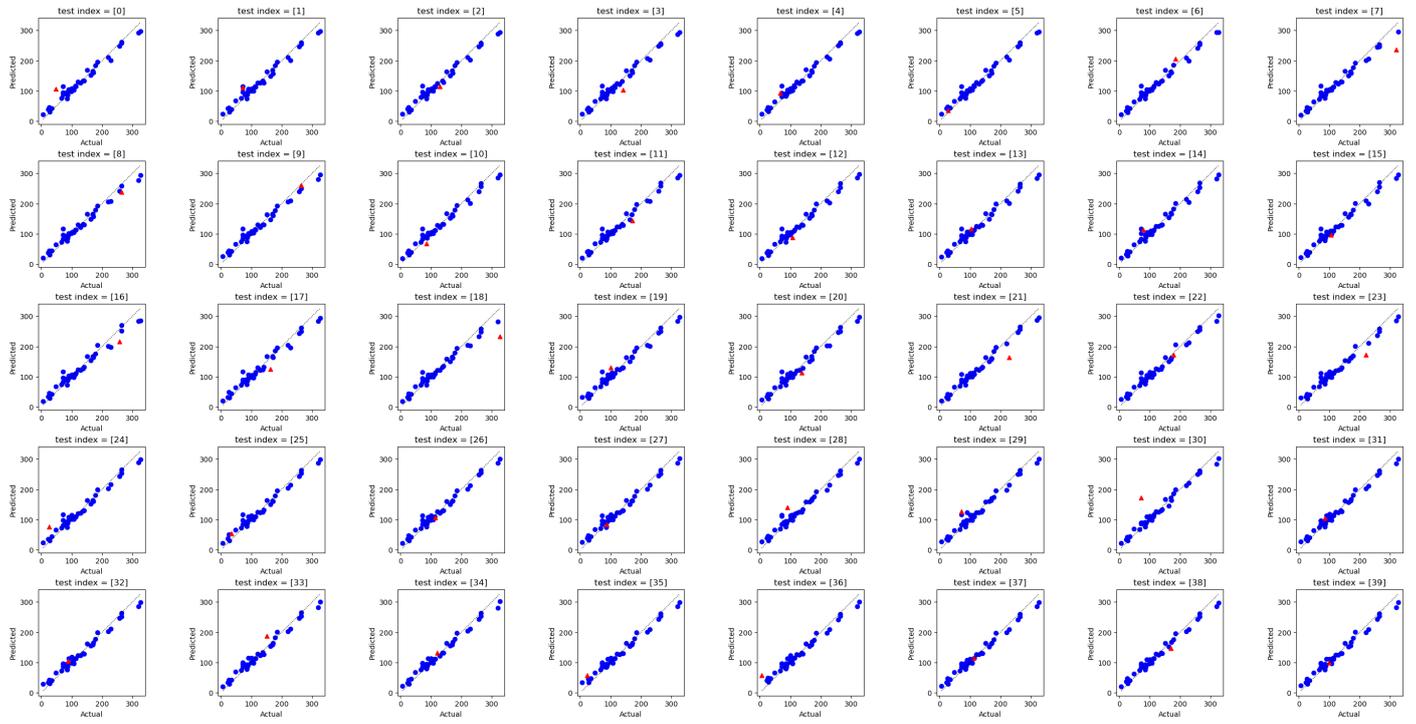

Figure S7. Actual vs. Predicted plots for all samples in Leave-one-out cross-validation. The target variable is fracture strength, with blue points representing the training data and red points representing the test data.

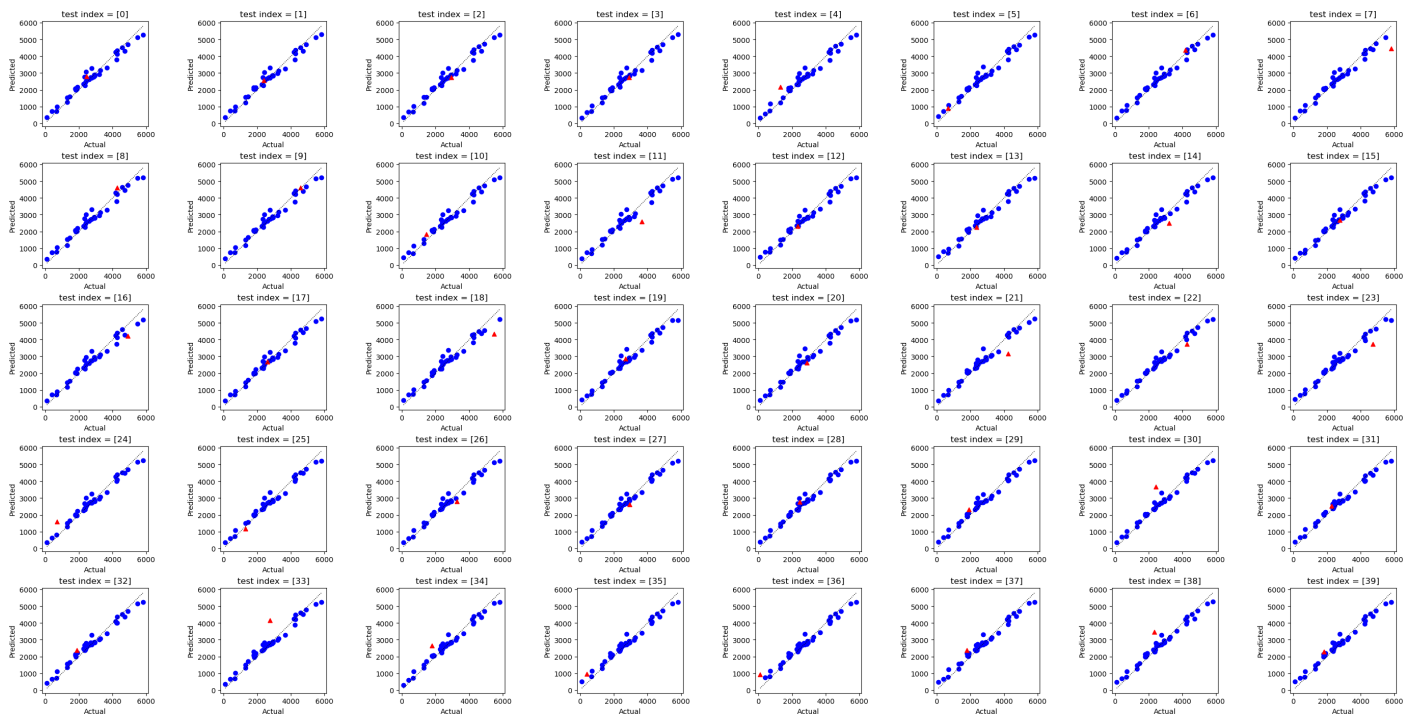

Figure S8. Actual vs. Predicted plots for all samples in Leave-one-out cross-validation. The target variable is electrical conductivity, with blue points representing the training data and red points representing the test data.

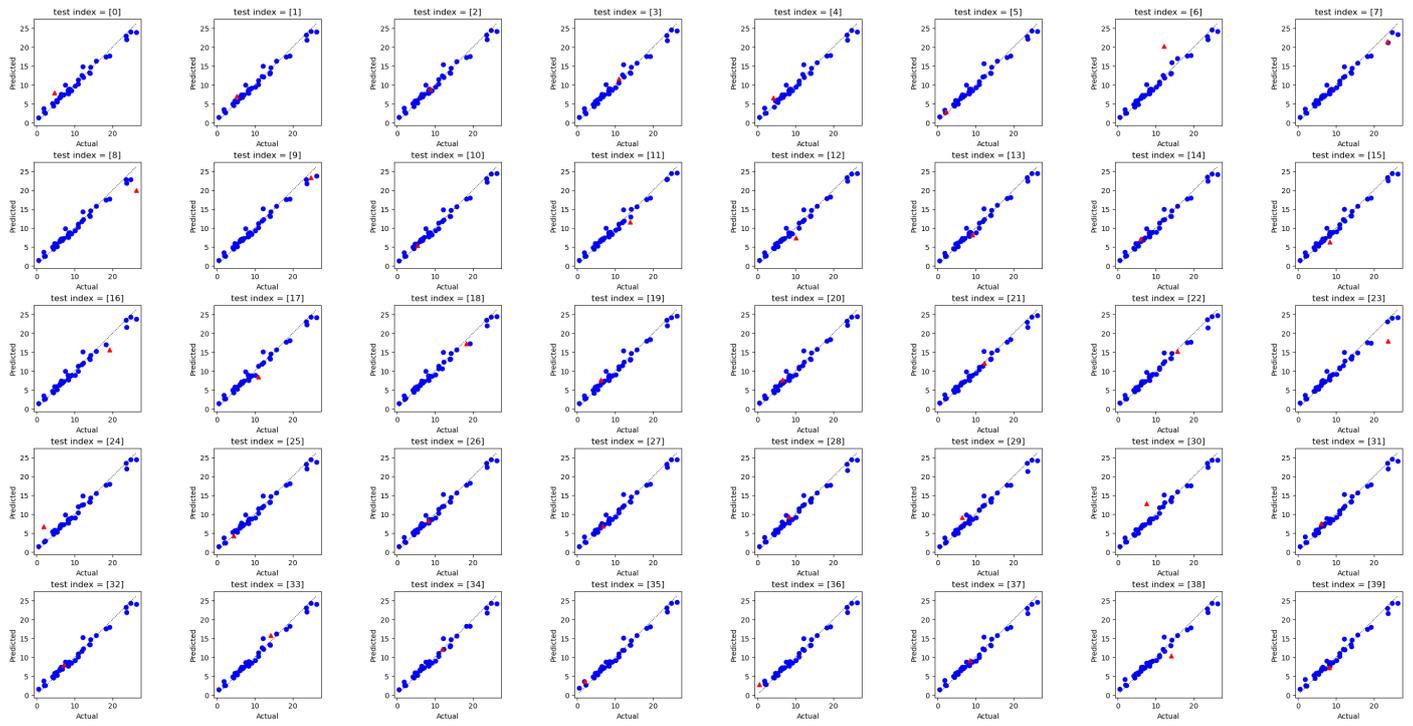

Figure S9. Actual vs. Predicted plots for all samples in Leave-one-out cross-validation. The target variable is Young's modulus, with blue points representing the training data and red points representing the test data.

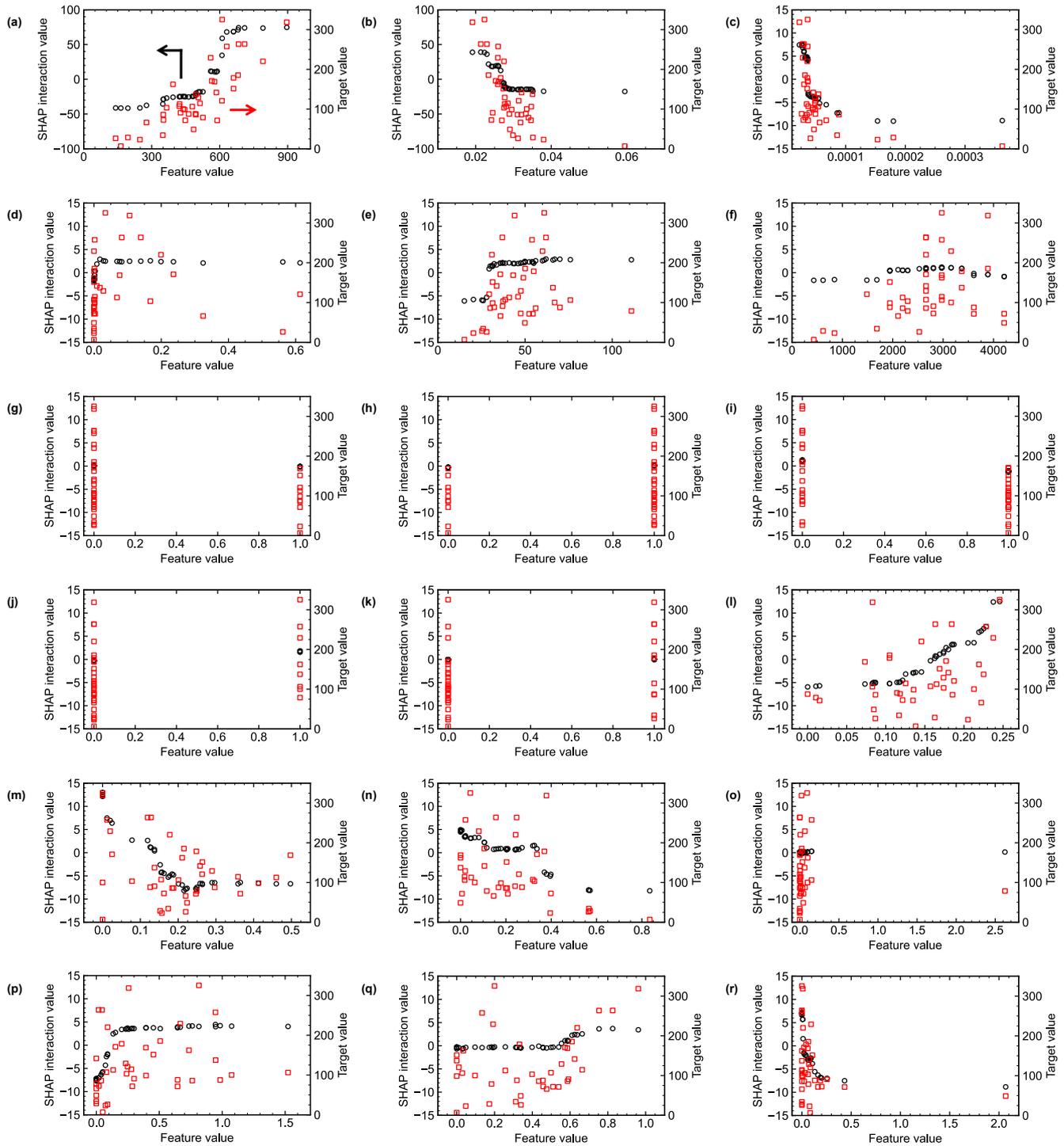

Figure S10. Relationship between feature values (the horizontal axis), SHAP interaction values for fracture strength [MPa] (the left vertical axis), and fracture strength [MPa] (the right vertical axis). The feature are (a) density [mg/cm], (b) linear density [mg/cm], (c) Cross-sectional area [cm$^2$], (d) void ratio, (e) $I_G / I_D$, (f) effective length, (g) dispersant: SC, (h) dispersant: TDOC, (i) dispersion method: Homogenizer, (j) dispersion method: Millser, (k) dispersion method: Nanovater, (l) DCS: Basis000, (m) DCS: Basis001, (n) DCS: Basis002, (o) DCS: Basis003, (p) DCS: Basis004, (q) DCS: Basis005, and (r) DCS: Basis006.

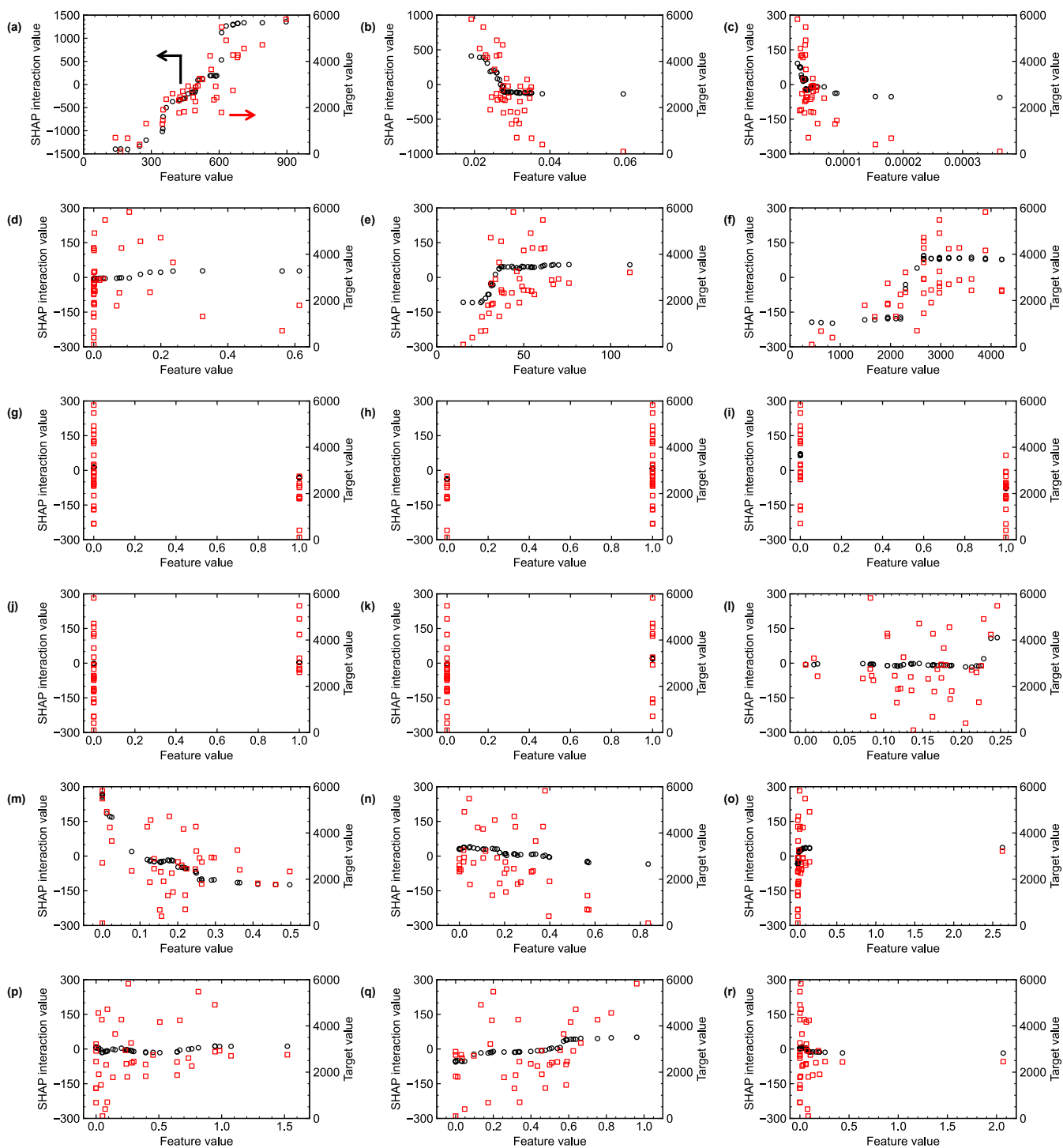

Figure S11. Relationship between feature values (the horizontal axis), SHAP interaction values for electrical conductivity [S/cm] (the left vertical axis), and electrical conductivity [S/cm] (the right vertical axis). The feature are (a) density [mg/cm], (b) linear density [mg/cm], (c) Cross-sectional area [cm$^2$], (d) void ratio, (e) $I_G / I_D$, (f) effective length, (g) dispersant: SC, (h) dispersant: TDOC, (i) dispersion method: Homogenizer, (j) dispersion method: Millser, (k) dispersion method: Nanovater, (l) DCS: Basis000, (m) DCS: Basis001, (n) DCS: Basis002, (o) DCS: Basis003, (p) DCS: Basis004, (q) DCS: Basis005, and (r) DCS: Basis006.

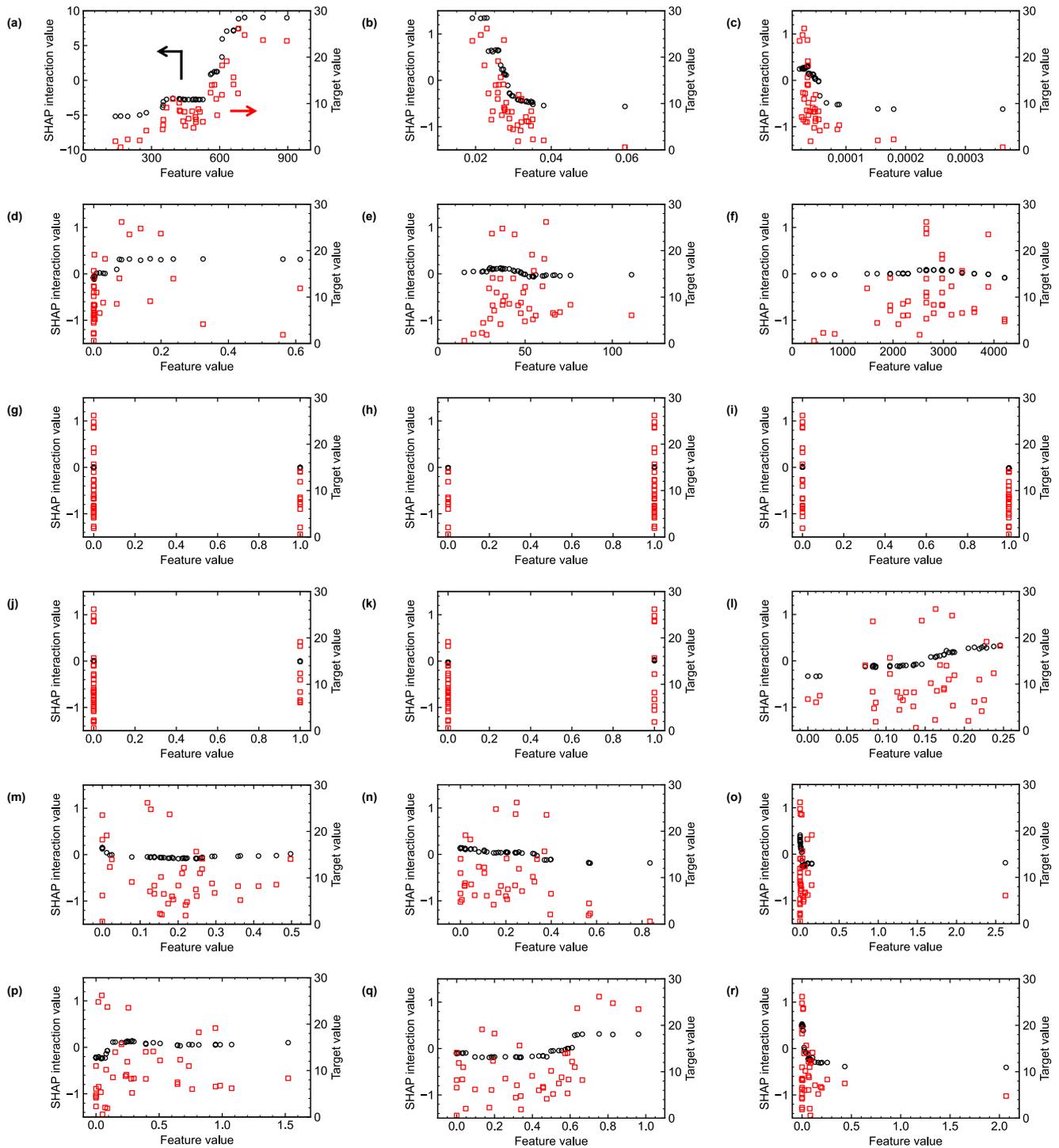

Figure S12. Relationship between feature values (the horizontal axis), SHAP interaction values for Young's modulus [GPa] (the left vertical axis), and Young's modulus [GPa] (the right vertical axis). The feature are (a) density [mg/cm], (b) linear density [mg/cm], (c) Cross-sectional area [cm$^2$], (d) void ratio, (e) $I_G/I_D$, (f) effective length, (g) dispersant: SC, (h) dispersant: TDOC, (i) dispersion method: Homogenizer, (j) dispersion method: Millser, (k) dispersion method: Nanovater, (l) DCS: Basis000, (m) DCS: Basis001, (n) DCS: Basis002, (o) DCS: Basis003, (p) DCS: Basis004, (q) DCS: Basis005, and (r) DCS: Basis006.